Review

# Solar Particle Acceleration

**Donald V. Reames** (https://orcid.org/0000-0001-9048-822X )
Institute for Physical Science and Technology, University of Maryland, College Park, MD, USA 20742

**Abstract** High-energy particles may be accelerated widely in stellar coronae; probably by the same processes we find in the Sun.  Here, we have learned of two physical mechanisms that dominate the acceleration of solar energetic particles (SEPs).  The highest energies and intensities are produced in "gradual" events at shock waves driven from the Sun by fast, wide coronal mass ejections (CMEs).  Smaller, but more numerous, "impulsive" events with unusual particle composition are produced during magnetic reconnection in solar jets and flares.  Jets provide open magnetic field lines where SEPs escape; closed magnetic loops contain this energy to produce bright, hot flares, perhaps even contributing to heating the low corona in profuse nanoflares.  Streaming protons amplify Alfvén waves upstream of the shocks.  These waves scatter and trap SEPs and, in large events, modify the element abundances and flatten the low-energy spectra upstream.  Shocks also reaccelerate residual ions from earlier impulsive events, when available, that characteristically dominate the energetic heavy-ion abundances.  The large CME-driven shock waves develop an extremely wide longitude span, filling much of the inner heliosphere with energetic particles.

# 1 Introduction

High-energy particles are likely accelerated in the coronae of most stars, but we can only study them directly near the Sun. Energetic particle radiation constitutes a recurring hazard to modern astronauts and missions in space and may have influenced the evolution of life here and on exoplanets around other stars.

Particle acceleration can only occur where the density is low enough that particles do not immediately dissipate their energy in Coulomb collisions. We have found that the most powerful physical processes for acceleration of solar energetic particles (SEPs) involve shock waves driven out from the Sun by fast, wide coronal mass ejections (CMEs) [1, 2], but regions of magnetic reconnection and turbulence in solar jets also produce electron-rich bursts of SEPs with unusual and distinctive element abundances although at lower average energies and intensities.

Energetic particles can be accelerated in and above the corona where the plasma densities fall rapidly below about $10^9$ atoms cm$^{-3}$; they rapidly lose energy in Coulomb scattering and stop at the higher densities of the low corona and chromosphere. The magnetic energy of reconnection can spawn energetic particles in the corona or generate heat below. In solar flares, accelerated particles trapped on closed magnetic loops soon scatter into the loss cone and plunge into denser plasma below to produce heat, X-rays, gamma-rays, and thermal EUV emission. Particles accelerated on open field lines in solar jets, called *impulsive* SEPs, escape along field lines where the streaming electrons generate type III radio emission. A signature of these impulsive SEPs is their unusual abundances including 10,000-fold enhancements of $^3$He/$^4$He and a systematic enhancement of increasingly heavy elements continuing as high as Pb [3 – 14]. In contrast, fast (> >500 km s$^{-1}$) CMEs drive shock waves that accelerate particles in *gradual* SEP events [1] that include the widespread, intense, energetic, and long-lasting SEP events [15 – 20].

# 2 Magnetic Reconnection

Since the Sun is composed of ionized plasma, differences in forces on ions and electrons, especially in differential rotation, cause currents that generate magnetic fields. Magnetic fields begin deep in the Sun but are deformed at the *tachocline*, the boundary between the inner radiative zone, which rotates uniformly, and the convective zone, which rotates faster at the equator than at the poles. As the plasma swirls, magnetically active regions or sunspots are visible at the radiatively-cooled photosphere [21]. When the fields become twisted and tangled, oppositely directed fields are brought together to cancel and reconnect, releasing energy. The magnetic energy that vanishes during reconnection drives all the activity on the Sun that we see as flares, CMEs, SEPs, X-rays, gamma-rays, and radio emission. Energy trapped in closed field configurations is emitted eventually as photons radiated from flares or can help drive out CMEs.

Patterns of field reconnection can be complicated, but Figure 1 shows two examples of topology that results when reconnection occurs either between open and closed lines, which allows SEPs to escape, or among completely closed lines, where they cannot. In Figure 1a, emerging loops have opposite polarity from the overlying field lines, with the energy in islands of reconnection driving SEP acceleration in the collapsing current sheet [22 – 24], thus producing a solar jet. In Figure 1a, newly opened field lines on the right are matched by newly closed fields on the left which capture some SEPs. These will eventually plunge into the denser corona to produce a hot solar flare. Thus, jets may always have an accompanying flare, but the SEPs we see do not come from this flaring region. This is why SEP properties were historically found to correlate poorly with associated X-ray intensities, for example [25, 26].

Figure 1b shows a topology where shear in the footpoints of an arcade of loops produces reconnection where all the field lines remain closed. All accelerated particles are trapped and eventually plunge into the footpoints of the loops to produce a flare with no SEPs in space [28]. However, the liberated flux rope shown in the figure may rise as part of a CME which can be fast enough to form a shock wave and produce a gradual SEP event. Note that the right-most cartoon in Figure 1 is often viewed end-on and the flux rope is shown as a completely open region (e.g. the Kopp and Pneuman model [27]). Other cartoons show it as a closed circle, disconnected from the Sun. In the likely topology, all particles accelerated in this reconnecting region plunge into the Sun to produce flaring.

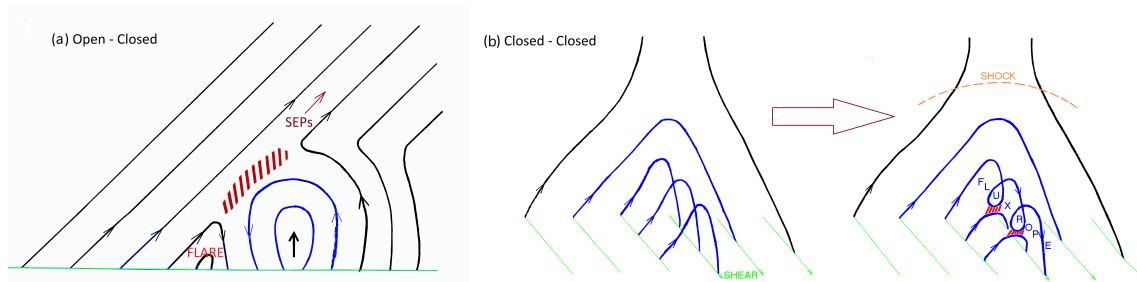

**Figure 1** Examples of the topology of magnetic reconnection are shown: (**a**) In a solar jet, initially open field lines of one polarity (*black*) reconnect with newly emerging loops (*blue*) in the *red hashed* region where accelerated SEPs can escape easily, however, some of the SEPs are captured in a newly closed region on the left, where they produce a hot flare. (**b**) A sheared arcade of field lined can reconnect in the *red hatched* regions to produce a semi-detached flux rope that can rise to form a CME. Here, all SEPs are trapped on closed field lines and eventually plunge into the denser corona to produce a flare. SEPs are accelerated in collisionless plasma; flaring is produced in collisional plasma [28].

Figure 1 does not cover all possibilities. If the reconnection region is too deep in the corona, there are no SEPs and the magnetic energy is dissipated as heat. Modifications of the topology in Figure 1a may also produce a CME and that CME may drive a shock wave that can reaccelerate the SEPs from the reconnection region. We will consider this possibility below. Of course, jets can be more complicated than the one shown [29 – 31].

# 3 Impulsive Events

Impulsive events were first associated with the streaming electrons that produce type III radio bursts, as distinct from the shock-accelerated electrons producing type II bursts [32, 33]. Later, some small SEP events were found to have 10,000-fold enhancements of $^3$He/$^4$He > 1 [34], and these electron- and $^3$He-rich events were subsequently found to be the same [35, 36]. It became clear that $^3$He enhancements were a resonance phenomenon produced by waves near the $^3$He gyrofrequency = $Q\omega_p/A$, where $\omega_p$ is the proton gyrofrequency, and the charge-to-mass ratio $Q/A = 2/3$ for $^3$He. The wave spectrum could be damped at the gyrofrequencies of the dominant ions, H at $Q/A = 1$, and $^4$He at $Q/A = 2/4$, but the low abundance of plasma $^3$He ($^3$He/$^4$He ≈ 5 ×10$^{-4}$) is unable to damp the resonant waves at that gyrofrequency. Many wave modes were suggested for the selective preheating of $^3$He [37-44], however, Temerin and Roth [45] suggested a more complete model with electromagnetic ion-cyclotron waves being generated by the streaming electrons that could be resonantly absorbed by mirroring $^3$He ions to accelerate them.  As acceleration increases, available $^3$He can become depleted completely [46], which contributes to spectral differences between $^3$He and $^4$He [47 – 49].

Subsequent observations also showed heavy-ion enhancements up to Fe [50, 10] that seemed to increase as $A/Q$ [8].  Later observations extended this pattern across the periodic table of elements, seen up to about Pb [4, 9, 11, 12].  This behavior was unlike the sharp resonance for $^3$He and was first associated with cascading waves [51 – 53] and later associated with a shock-like acceleration as ions are scattered back and forth from the ends of collapsing islands of magnetic reconnection [22].

# 4 Shock Acceleration

The dominant mechanism for producing high-energy particles is acceleration by shock waves [54] driven out by the Sun by fast, wide CMEs with speeds of ~800 – 3000 km s$^{-1}$. When the magnetic field is quasi-parallel to the shock normal, particles can be scattered back and forth across the shock by resonant scattering against waves upstream and downstream [55, 56, 15, 16, 57].  As the particles stream away from the shock at each higher energy, they increase the scattering by amplifying resonant waves [58, 59] of wave number $k$ with

$$k \approx B/\mu P \qquad (1)$$

where $P$ in the magnetic rigidity, i.e. momentum per unit charge, of the particle and $\mu$ is its pitch angle with respect to the magnetic field of vector $\boldsymbol{B}$.   On each transit of the shock, particles gain an increment in velocity related to the plasma-wave velocity difference across the shock.  The growth of self-amplified waves is an essential aspect of shock acceleration.  For quasi-perpendicular regions of the shock, where the magnetic field lies nearly in the shock plane, particles are also accelerated in the $V_{shock} \times \boldsymbol{B}$ electric field [60].  Some early authors [15] assumed $\mu \approx 1$, so each rigidity had its own unique resonant wave number, a convenient simplification.

For even greater simplicity, many authors assume that particle scattering does not vary at all with time. More generally, Bell [56, 57] assumed that equilibrium is maintained between particles and resonant waves and this assumption was applied to SEP shocks by Lee [15, 16].  Later Ng and Reames [57] allowed full time evolution of both particles and waves, allowing estimates of the timescale of shock acceleration.

Element abundances accelerated by shock waves tend to reflect the abundances in the corona at 2 – 3 solar radii [61 – 63] where the ions are first sampled.  The coronal abundances differ from abundances in the photosphere as a function of the first ionization potential (FIP) because of ion-neutral fractionation as the particles cross the chromosphere to the corona [64, 65] where they all become ionized.  Averaged SEP abundances can be a good measure of coronal abundances [66, 67] although their FIP pattern differs from that seen in the solar wind [68 – 71].  SEPs are *not* just accelerated solar wind.

For impulsive SEP events, we were able to neglect scattering of the ions during transport since impulsive events have low intensities, the particles tend to propagate scatter free [72], and they are focused by the diverging magnetic field.  However, wave scattering is a basis of shock acceleration since the particles streaming away grow waves resonant with higher- and higher-energy ions during acceleration, and these waves extend farther and farther from the shock as intensities increase [16, 57]. The wave intensity and thus particle scattering is a strong function of space and time. For different ion species of a given velocity, the rigidity dependence of scattering tends to trap low-rigidity (low-$A/Q$) ions near the shock while the higher-rigidity (higher-$A/Q$) ions leak away. This power-law dependence upon $A/Q$ was first directly observed by Breneman and Stone [73]. Thus, early in an event, Fe/O will be first observed to be enhanced, for example, although later, when the shock approaches, Fe/O will always be depressed from its coronal value. With the proper averaging we should be able to recover the source coronal abundance. In small gradual events the depletion in Fe/O, from Fe loss by preferential leakage, will be most noticeable throughout. In large gradual events, preferential trapping of O near the shock can cause elevated Fe/O to persist for a day or so at moderate energies until the shock actually passes over the observer [7].

During the early acceleration, as intensities of protons of a given rigidity rise at a shock, they first stream away, causing more and more resonant waves to grow. Wave scattering reduces this streaming, until an equilibrium limit is reached where increasing intensities at the shock just grow enough additional waves to stop the streaming out to observers away from the shock. This has been called the "streaming limit" that causes an intensity plateau [74 – 76], or the "flat spectra" rediscovered recently [77]. Figure 2a shows superposed time profiles for several large SEP events where the early proton intensities near 1 AU are bounded at just over 200 protons cm$^{-2}$ sr$^{-1}$ s$^{-1}$ MeV$^{-1}$ although in some events intensities rise to much higher values later, near the shock [74]. If source intensities increase, the protons streaming outward simply grow more resonant waves so as to restrict the streaming, trapping ions near the source. Figure 2b shows flattening of plateau H and O spectra, typical of all species, for 5 SEP events [76]. Figure 2c shows the importance of high intensities at ~10 MeV for flattening the intensities at energies below [78 – 80]. Ng [81] discusses the way the limiting proton intensities vary with radius and with plasma and shock parameters.

The streaming limit is a common feature in all large gradual SEP events with sufficiently high intensities. For stronger shocks, the flattening extends to higher energies. It is a transport-generated limit involving equilibrium between streaming protons and resonant waves. When there is no streaming, e.g. at the shock, there is no intensity limit. Also, when intensities rise extremely rapidly, they can briefly exceed the limit, until enough waves grow to reestablish it. This situation occurs in the Jan 20 2005 event, as shown in [80], and is well-described by theories that follow the time evolution of both particles and waves [79].

The streaming limit decouples particle intensities near the shock from those farther upstream, creating intensity-peak structure near the shock, historically called the energetic storm particle (ESP) event. Following Bell [55, 56] and Lee [15, 16], Vainio et al. [82] have produced a model of this ESP structure in the foreshock region. As the ESP event moves away from the Sun the average field intensity decreases and waves of given $k$ tend to resonate with particles of lower rigidity, allowing the highest-energy particles to leak away first. However, in very large events continued acceleration can slow this leakage.

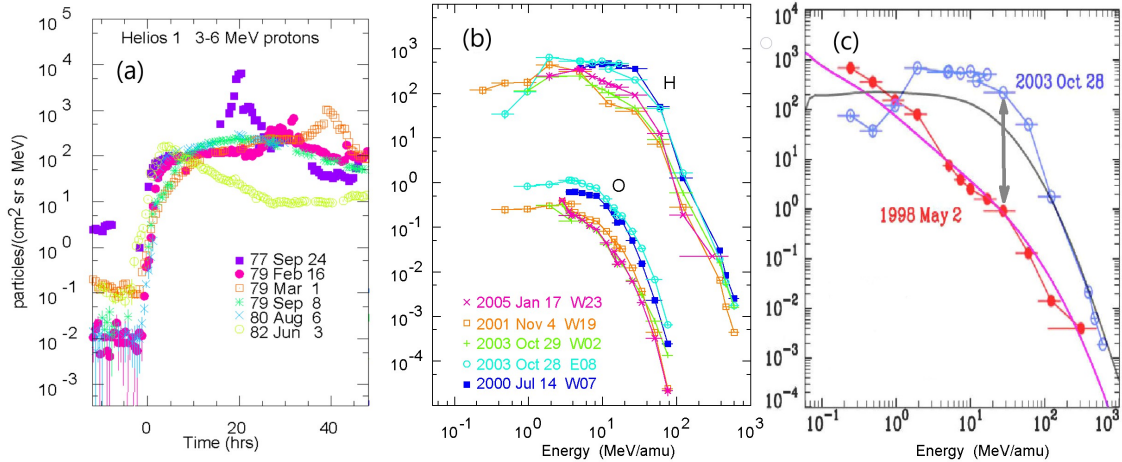

**Figure 2** (**a**) Superposed time profiles of 3 – 6 MeV protons show limited plateau intensities early in large SEP events that often rise later when the shock approaches [74]. (**b**) H and O plateau spectra for several events show flattening at low energies [76]. (**c**) Theoretical fits are compared, using the Ng et al. [79] theory, for the flattened plateau proton spectra of the large 2003 October 28 event and the un-flattened, power-law spectrum in the smaller event of 1998 May 2 [78]. The ~100-fold intensity difference above ~10 MeV is emphasized by the gray arrow; the higher intensities amplify more waves that restrict outward flows, trapping ions and flattening spectra.

## *4.1 Spatial Distributions*

Spatial distributions of SEPs driven by shock acceleration can be quite extensive. The earliest particles are released at ~2 solar radii [61, 62] and particles from these earliest spectra are effectively the dominant seed population for all additional acceleration. As the shock moves outward wave turbulence tends to trap particles with the shock, but $B$ decreases on average, so the existing waves tend to resonate with ions of lower rigidity (Eq. 1), allowing the highest rigidity ions to leak away first.

The earliest particles that escape the shock when it is near the Sun usually contain energies of greatest radiation hazard. Figure 2b shows early proton spectra up to 700 MeV and some are flat out to ~50 MeV. A recent measurement near the Sun at 0.35 AU [23] also observed the flat spectra but only below ~1 MeV. Figure 3 shows measurements by spacecraft conveniently located to show the radial evolution of the central region in a gradual SEP event. *Helios 1* is well connected to the center of the shock near the Sun, *Helios 2* and IMP 8 when it is near 1 AU, and *Voyager 2* when it is near 2 AU, as shown in Figure 3b. Proton intensities at 6 – 11 MeV are compared in Figure 3a and more complete energy coverage is shown in Figures 3c – 3e. Protons of 100 – 200 MeV peak early, are still visible at 1 AU, but have dissipated by 2 AU. In contrast 20 – 40 MeV protons still peak at the shock near 2 AU in this event.

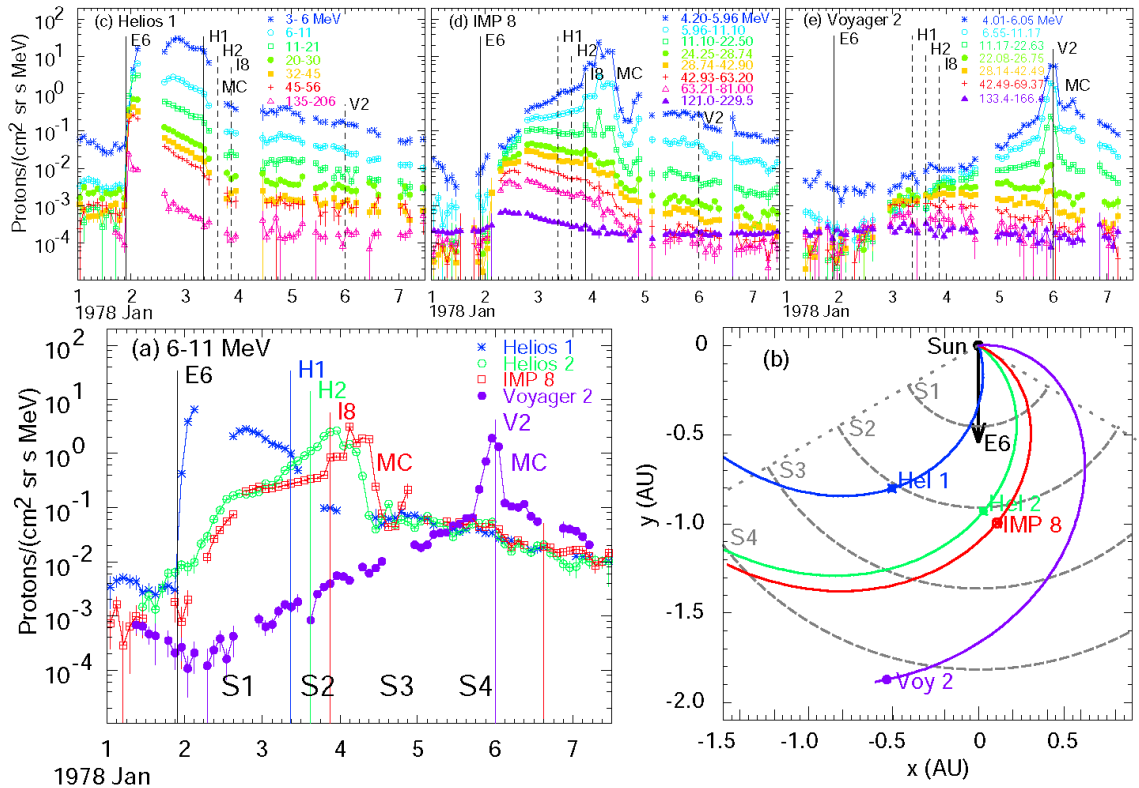

**Figure 3** (**a**) The radial evolution of 6 – 11 MeV proton intensities is captured by four spacecraft as the accelerating shock wave moves out to ~2 AU in the 1978 Jan 1 SEP event. (**b**) The spatial distribution of the spacecraft is shown and four shock positions S1 – S4 correlate with times in (**a**). Full energy coverage vs. time from ~3 – 200 MeV is shown for (**c**) *Helios 1*, (**d**) IMP 8, and (**e**) *Voyager 2* [18, 83].

While the *Helios*-IMP8-*Voyager* period provided several revealing coincidences including a case of successive strikes of field lines by both flanks of a wide shock in September 1978 [18, 83], the STEREO period showed the wide distribution of SEPs and of the shock waves that produce them. The example in Figure 4 shows intensities of 20 MeV protons at the widely-spaced spacecraft distribution shown in Figure 4b. The shock itself is seen at all three spacecraft at similar times, suggesting a nearly spherical shape.

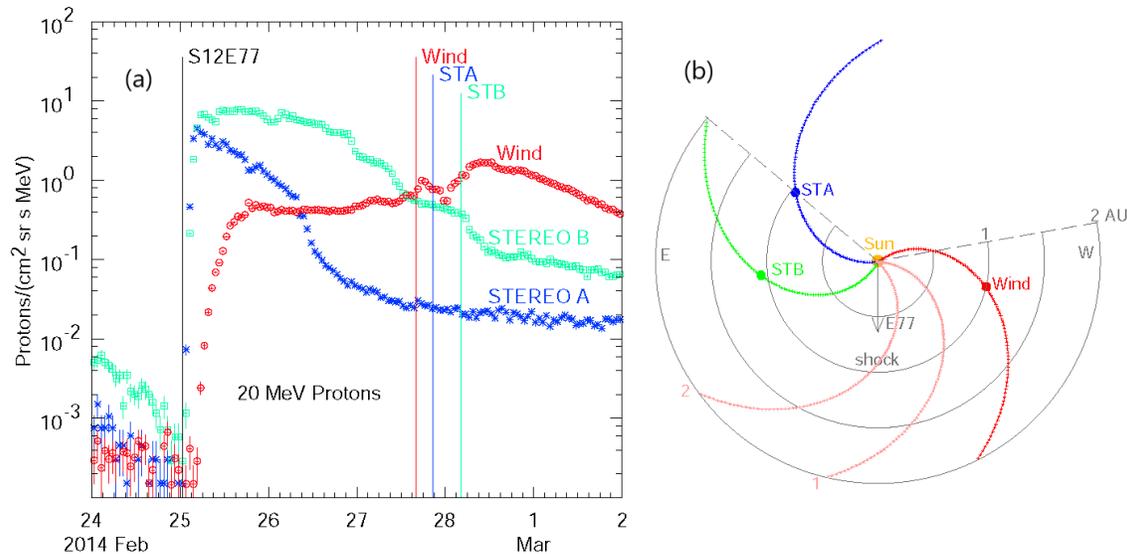

**Figure 4 (a)** Intensities of 20 MeV protons vs. time are shown for three spacecraft located as in (**b**). Similar shock arrival times at the spacecraft indicated in (**a**) suggest a nearly spherical shock [82]. The pink field lines labeled 1 and 2 in (**b**) will rotate to intercept *Wind* when the shock reaches 1 or 2 AU, respectively, bearing some of the SEPs acquired at their initial longitudes.

## 4.2 Element Abundances and Seed Particles

Abundances of elements, accelerated by shock waves to a given velocity, or energy per nucleon, tend to reflect the abundances of the seed particles initially injected into the shock. In most gradual SEP events this is the ambient abundances of the solar corona near the active region where the shock begins, but shocks will accelerate any (and all) seed particles that are available. The question is: which will dominate?

Mason, Mazur, and Dwyer [84] first noticed that shocks could reaccelerate $^3$He and other impulsive ions provided by earlier impulsive SEP events. It was recognized [85 – 88] that quasi-perpendicular shocks might selectively prefer the pre-accelerated seed population of ions that had sufficient speed to overtake the shock from behind, i.e. residual suprathermal particles from impulsive events. It turns out that residual ions from many persistent impulsive flares can accumulate above active regions, providing a seed population, in addition to the ambient corona plasma, thus complicating the abundance pattern of elements in SEP events. Reames [89] identified four processes leading to different SEP abundance patterns:

SEP1: "Pure" impulsive SEPs with enhanced heavy elements from reconnection in solar jets, but no fast shocks.

SEP2: Impulsive SEP1 ions reaccelerated as seed particles of a fast CME-driven shock from the same jet [90 – 92] dominate high $Z$.

SEP3: Enhanced abundances accelerated by a fast CME-driven shock traversing an active region with residual seed ions from many small impulsive events that will dominate high $Z$.

SEP4: Particles accelerated by a fast, wide, CME-driven shock with only ambient coronal abundances available as the dominant seed particles.

Since shock waves will accelerate any ions available, SEP2 and SEP3 events tend to reflect both types of seed particles, the heavy-ion-rich impulsive seeds dominate the high-Z SEPs, and the ambient abundances are left to dominate protons. These patterns are compared in Figure 5. Fits to the abundance enhancements vs. $A/Q$ involve varying plasma temperatures to determine ion $A/Q$ values from their well-determined temperature dependence [93, 94]. Best-fit temperatures are shown in Figure 4b and fit lines in Figure 4c, for three time-intervals. The events in the left panel in Figure 4c show rising enhancements vs. $A/Q$ for shockless SEP1 impulsive SEP events, the event in the right panel shows SEP4 falling enhancements vs. $A/Q$ where higher-$A/Q$ elements from the ambient-coronal seeds have preferentially leaked away, and the middle panel shows a SEP2 or SEP3 event with components of both rising and falling seed populations.

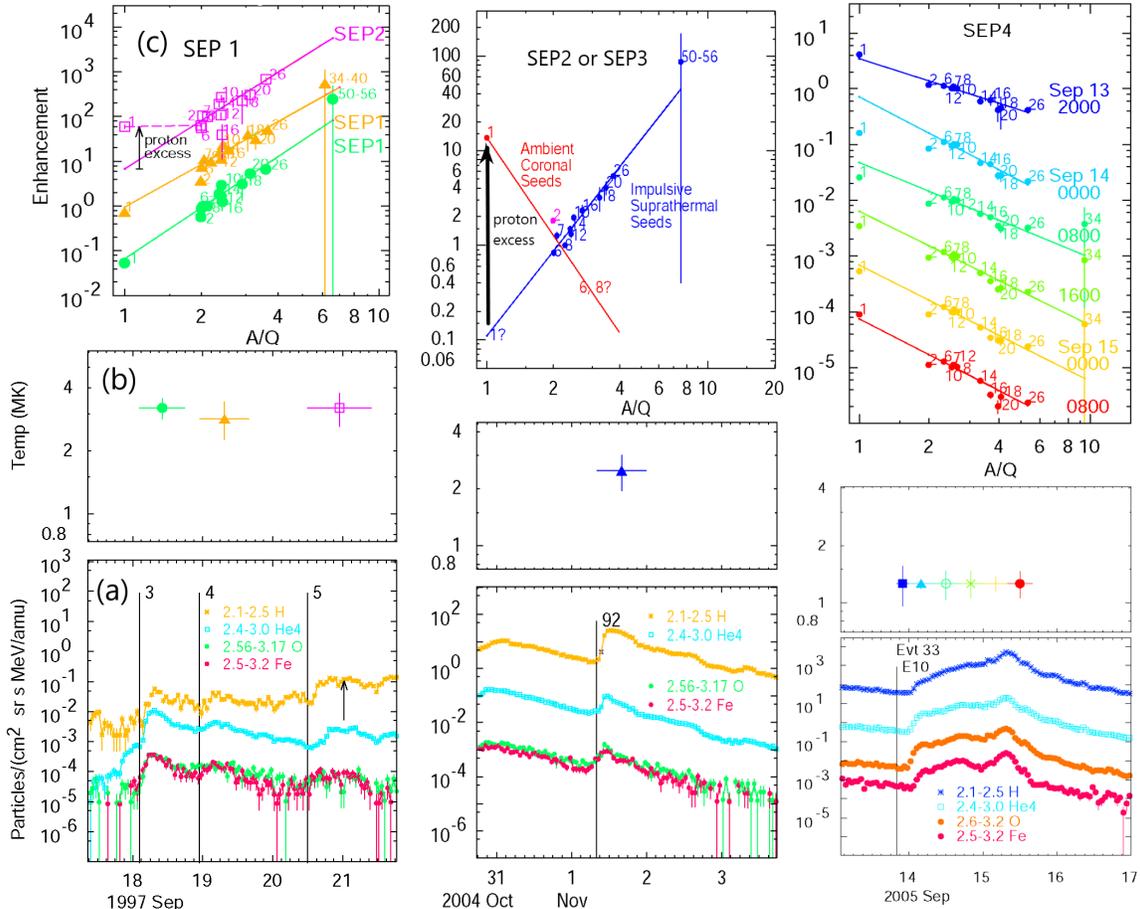

**Figure 5** In (**a**) intensities of H, $^4$He, O, and Fe in the listed intervals of MeV amu$^{-1}$ are shown for three time intervals. (**b**) Temperatures derived from best-fits of enhancement vs. $A/Q$ show with power-law fits in the same color in (**c**) above [89 95, 96].

In the impulsive SEP1 events, protons either lie on the fit line, or perhaps, theory allows the increase to begin above He [22] so H and $^4$He would both be equally unenhanced. For SEP4 events protons again lie on the fit line since all ions belong to a single seed-particle population. For SEP2 and SEP3 events, proton enhancements signal the presence of two seed populations. Sequences of SEP3 events, dominated by impulsive seed ions, can come from a single active region, [97]. These help to distinguish SEP3 events from smaller, isolated SEP2 events.

Very intense SEP4 events do have rising initial $A/Q$ dependence when only the higher-$Z$ ions can penetrate out to the observer. For these SEP4 events, the $A/Q$- dependence suddenly declines as the shock passes; for SEP3 events, the enhancements persist across the shock since they are a fundamental property of the impulsive seeds, not of transport. Element abundances and their time dependence is a powerful tool in the study of SEP acceleration and transport.

Using the assumption of power-law abundance enhancements to derive source plasma temperatures may seem uncertain, but Bučík et al. [98] have now directly confirmed similar temperatures of ~2.5 MK in the source jets that produce impulsive SEP events. Also, nearly all of the SEP abundance studies have used Maxwellian electron distributions to determine $A/Q$ vs. temperature. However, using kappa distributions for the thermal electrons, Lee et al. [99] have shown that the Maxwellian distributions used to derive $A/Q$ vs. temperature [93, 94] did not produce large errors. They also suggested that since the SEP and source temperatures are similar, the SEPs must leave the acceleration region on a time scale shorter than their ionization time. After acceleration these ions evidently pass through enough matter to strip them to an equilibrium ionization state that varies with velocity [100] and leaves $Q_{Fe} \approx 20$ near 1 MeV amu$^{-1}$. This stripping suggests [100] that impulsive SEP events occur at about 1.5 solar radii. These ions from SEP1 events must retain their higher "stripped" charge states if they get reaccelerated by shock waves to produce SEP2 or SEP3 events, even though the shock acceleration may occur at the much lower densities at $2 - 3$ solar radii. Thus, the events with impulsive seeds have enhanced Fe/O and charge states appropriate to their stripping energy *after* the SEP1 phase, but their abundance-derived temperatures and $Q$-values are appropriate to their earlier state before and *during* the SEP1 phase. In contrast, the SEP4 events have ambient coronal charge states (at ~ 1 MK) with coronal source abundances modified only by transport.

## *4.2 Energetic Storm Particle (ESP) Events*

The ESP structure, of particles trapped by self-amplified waves near a strong shock, forms early when the shock is near the Sun [16, 82], but the field decreasing with time allows the highest energies leak away as the shock travels out to 1 AU. Nevertheless, it is important to recognize that ESP events may still contain significant intensities of >100 MeV protons at 1 AU as seen in examples shown in Figure 6. An ESP spike at the shock can even be observed in >700 MeV protons in the 2001 November 6 ESP event (see Figure 5.1 in [2]).

Kouloumvakos et al. [101] have used the ESP model of Vainio et al. [82] as a source function at the shock to model the multi-spacecraft spatial distribution and onset timing of SEPs. They find that cross-field diffusion is completely unnecessary for their model. Finally, a shock model may have completely replaced the old shockless "flare plus perpendicular-diffusion" models that once prevailed [2, 105].

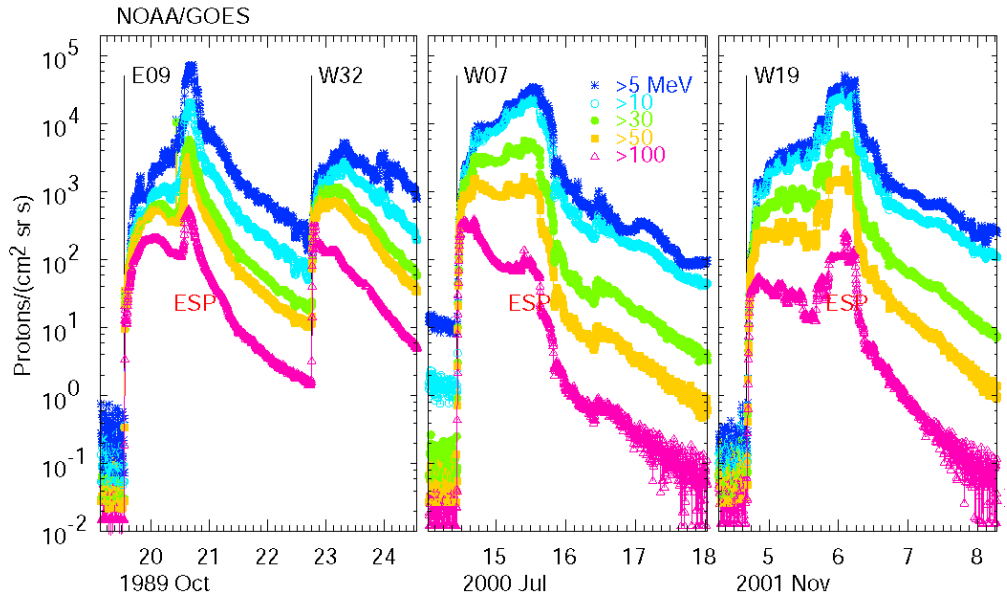

**Figure 6** For large SEP events near central meridian, intensities of high-energy protons trapped near the shock can sometimes exceed those released early, even at energies > 100 MeV as seen in samples here.

## 5 Discussion

Recently, Mason et al. [102] suggested that the enhancement of Fe/O could come from the suppression of O which resonates with waves damped by $^4$He at $A/Q = 2$. However, this fails to explain the strong continuing enhancement of heavier elements well above Fe, fairly well approximated as a power law in $A/Q$ observed at all energies [9, 11,12]. This power-law increase does not exclude the possibility that additional resonant wave absorption can produce selective peaks in Si and S in some small events [103], presumable near $A/Q \sim 3.0$ as the second harmonic of $^3$He at $A/Q = 1.5$. Turbulent reconnection in jets may well produce both power-law reconnection [22] and resonant [104] processes, especially in the small SEP events with steep spectra that have only been seen below ~1 MeV amu$^{-1}$. Jets that produce larger impulsive SEP events with O and Fe spectra measurable out to 10 MeV amu$^{-1}$ or more are likely to include CMEs that can drive shock waves to 500 km s$^{-1}$ or more, reaccelerate the SEPs, and perhaps even smooth the $A/Q$ dependence by including extensive or multiple reconnection regions that average differing characteristics. In the list of Fe-rich events at 3 – 10 MeV amu$^{-1}$ presented by Reames et al. [9], about 40% of the events have CMEs with measured speeds above 500 km s$^{-1}$; these are surely SEP2 events and many of them are probably SEP3 events that average extensive sources of seeds. Generally, they are fit with power-laws in $A/Q$ determined at temperatures of 2.5 – 3.2 MK [6, 95]. Despite the dominance of power-law dependence, resonant wave absorption processes may still contribute the occasional additional enhancements of some elements.

For gradual SEP events, we have come a long way since Gosling's "Solar flare myth" [105, 106] pointed out the dominance of CMEs and shock waves, rather than flares, in interplanetary phenomena, including SEPs. The STEREO mission has shown how shock waves, and the SEPs they accelerate, can envelope the Sun (e.g. Figure 5) [83], while gamma-ray-line measurements suggest that the particles that are accelerated and magnetically trapped in flares are $^3$He-rich [107,108] and Fe-rich [109] like those of the impulsive SEPs we now know to come from jets [14]. Thus, impulsive SEPs are logically connected to flares, but do not actually come from flares; they come out to us from jets. For gradual events we must consider the hierarchy of seed particles, when available, and their reacceleration by shocks.

In most large gradual events, heavy ion abundances seem to be dominated by either impulsive (SEP3) or ambient-coronal (SEP4) seed ions across the whole longitude span of the shock [110, 96]. However, one might expect to find SEP events where the shock has sampled different ions at different longitudes. Recently, Xu et al. [111] have found just such an event, where spacecraft separated by ~60° longitude clearly see Fe-rich material at one spacecraft and Fe-poor material at the other over the full energy range of $0.1 - 10$ MeV amu$^{-1}$.

Resonant wave-particle interactions are an essential factor in SEP acceleration. They selectively enhance $^3$He in impulsive SEP events [39, 45], they enable the dominant acceleration as they scatter ions back and forth across a shock [15, 16, 57], and, as intensities increase, they produce rigidity-dependent scattering that can flatten spectra and enhance or suppress element abundances [112, 113, 79, 75, 76]. These abundances and spectra tell us about the underlying physical processes.

Parker [114] proposed the release of magnetic energy in current sheets of many small *nanoflares* as a mechanism for heating the entire solar corona. If this is the case, it is easy to imagine the acceleration of energetic particles in magnetic loops that carry that energy downward and deposit it in the footpoints of the loops. Electrons emit much of their energy as X-ray bremsstrahlung but stopping ions largely heat the plasma. Thus, SEP acceleration may actually be a significant agent in the heating of solar and stellar coronae.

For stars, other than the Sun, there is information on flares but not on CMEs or SEPs. Kahler's [115] paper on "big flare syndrome" provided the first caution that a correlation between SEPs and flare-related parameters, for example, does not mean that flares *cause* SEPs, but rather suggests that increasingly large dissipation of magnetic energy can result in many increasing phenomena. Nevertheless, big flare syndrome does suggest that flare stars may have more intense SEPs, so the underlying correlation is explored further by Kahler and Ling [116]. Other studies like Hu et al. [117] apply CME models to superflares, while Fu et al. [118] study the dependence of SEP properties on stellar rotation speeds.

Recently, Mohan et al. [119, 120] have detected 550–850 MHz radio emission similar to solar type IV radio emission from the young active M dwarf star AD Leo. Solar type IV radio bursts are produced by energetic electrons in post-flare loops and the moving flux ropes associated with fast (>900 km s$^{-1}$), wide (>60°) CMEs [121], the same type of CMEs known to produce large, gradual SEP events [1, 20]. The ability to detect and measure fast, wide CMEs could greatly increase our knowledge about large SEP events in other stellar systems.

# Conflict of Interests

The author declares that this research was conducted in the absence of any commercial or financial relationships that could be construed as a potential conflict of interest.

# Funding

No institutional funding was provided for this work.